\documentclass[twocolumn]{jpsj2}

\def\runauthor{Yuki \textsc{Fuseya} and Yoshikazu \textsc{Suzumura}}

\title{%
Superconductivity and Density Wave in the Quasi-One-Dimensional Systems: 
Renormalization Group Study
}

\author{%
Yuki \textsc{Fuseya}\thanks{E-mail address: fuseya@slab.phys.nagoya-u.ac.jp} and 
Yoshikazu \textsc{Suzumura}
}

\inst{%
Department of Physics, Nagoya University, Nagoya 464-8602
}

\recdate{\today}

\abst{%
	The anisotropic superconductivity and the density wave have been investigated by applying
	the Kadanoff-Wilson renormalization group technique to the quasi-one-dimensional system
	with finite-range interactions.
	It is found that a temperature ($T$) dependence of response functions is 
	proportional to $\exp(1/T)$
	in a wide region of temperature even within the one-loop approximation.
	Transition temperatures are calculated
	to obtain the phase diagram of the quasi-one-dimensional system, 
	which is compared with that of the pure-one-dimensional system.
	Next-nearest neighbor interactions ($V_2$) 
	induce large charge fluctuations, which suppress the $d_{x^2 -y^2}$-wave 
	singlet superconducting ($d$SS) state
	and enhance the $f$-wave triplet superconducting ($f$TS)
	state. 
	From this effect, the transition temperature of $f$TS becomes comparable
	to that of $d$SS for large $V_2$,
	so that field-induced $f$-wave triplet pairing could be possible.
	These features are discussed to comprehend the experiments on the (TMTSF)$_2$PF$_6$ salt.

}

\kword{%
quasi-one dimension, renormalization group, anisotropic superconductivity,
triplet superconductivity, (TMTSF)$_2$PF$_6$
}

\begin{document}

\maketitle

\section{Introduction}

	The mechanism of superconductivity in quasi-one-dimensional (q1d) systems
	has been attracting renewed interest since the appearance of
	the triplet superconductivity suggested in the
	experiments on the q1d organic conductor (TMTSF)$_2$PF$_6$.
	\cite{Hc2,Knight}
	In the phase diagram of (TMTSF)$_2$PF$_6$\cite{Jerome}, 
	the superconducting phase is adjacent to the spin density wave (SDW) phase.
	From the knowledge of the anisotropic superconductivity,\cite{MSV,SLH,Emery}
	which have been extensively enriched for two decades,
	the $d$-wave singlet superconductivity ($d$SS) is expected to occur 
	in such superconducting phase.
	However, recent experiments on (TMTSF)$_2$PF$_6$ 
	indicate a counter example of this widespread knowledge.
	The unsaturated magnitude of the upper critical field $H_{{\rm c}2}$\cite{Hc2}
	and the constant Knight shift thorough the superconducting transition temperature
	$T_{\rm c}$\cite{Knight}
	strongly suggest the formation of triplet pairing.
	In addition, the SDW phase turned out to be an exotic one,
	in which the $2k_F$ charge density wave (CDW) coexists 
	with the $2k_F$-SDW as seen from the X-ray experiments.\cite{PR,Kagoshima}

	The theoretical study of q1d systems
	is complicated due to its dimensionality originated from the anisotropy,
	where the effects of the one-dimensional (1d) fluctuation 
	are essential to understand its physics.
	In the system with the 1d fluctuation,
	there are 
	equally divergent
	contributions from both Cooper (electron-electron)
	and Peierls (electron-hole) channels.
	However, the 1d fluctuation 
	cannot be evaluated correctly by the conventional method 
	such as
	the random phase approximation (RPA)\cite{Shimahara},
	the fluctuation exchange (FLEX) approximation\cite{KK,KAA}
	or the third-order perturbation theory,\cite{NY}
	which
	are powerful for the study of three- or 
	two-dimensional anisotropic superconductivity.
	On the other hand, 
	the renormalization group (RG) technique based on the g-ology
	is extremely useful for the treatment of the 1d fluctuation,
	whereas it is complicated to deal with anisotropic superconducting states,
	such as $d$SS.

	Recently, Duprat and Bourbonnais extended the technique of Kadanoff-Wilson
	RG for purely 1d systems
	to that for q1d systems.\cite{DB}
	This technique enabled us to treat both 1d fluctuations and 
	various kinds of anisotropic 
	superconductivity.
	Meanwhile, the essential role of charge fluctuation has been pointed out 
	phenomenologically.\cite{KAA,FOKM}
	Quite recently, it is shown by the RPA\cite{TK} that 
	charge fluctuations induced by next-nearest repulsive interactions
	can give rise to the $f$-wave triplet superconductivity ($f$TS), 
	though the 1d fluctuation effects are still pushed aside.

	The purpose of the present paper is to get a deep insight into
	the interplay of the superconductivity and the density waves in q1d systems
	in terms of the Kadanoff-Wilson RG technique developed for q1d system.
	Especially, we shall focus on clarifying
	i) crossover from purely 1d case into q1d case,
	ii) effects of long-range Coulomb interactions
	and iii) microscopic mechanism of the q1d superconductivity,
	considering the effect of the 1d fluctuation.

	In \S 2, we briefly describe the Kadanoff-Wilson RG technique.
	Next, the method of the present work.
	In \S 3,
	it is shown that the response functions
	exhibit a noticeable property being proportional to $\exp (1/T)$.
	Phase diagrams for several magnitudes of {\it inter}chain hopping
	are demonstrated on the plane of {\it intra}chain interactions.
	Transition temperatures as a function of the {\it inter}chain hopping, 
	are also presented.
	Then a discussion of the mechanism of the q1d superconductivity
	is given.
	Section 4 is devoted to conclusion and 
	a comparison of the present result and the experiments.

\section{Formulation}
\subsection{Model}

	We consider a system consisting of an array of $N_b$ chains of length $L$,
	which is described by the Hamiltonian,
\begin{align}
	H &= \sum_{p, {\bf k}, \sigma}\xi ({\bf k})a^{\dagger}_{p{\bf k} \sigma}a_{p{\bf k} \sigma}
	 \nonumber \\
	& +\frac{\pi v_F}{LN_b}
	\sum_{{\bf k}_{i},{\bf q},\sigma}(g_2 
	\delta_{\sigma_1 \sigma_4}\delta_{\sigma_2 \sigma_3}
	-g_1 \delta_{\sigma_1 \sigma_3}\delta_{\sigma_2 \sigma_4}) \nonumber \\
	& \times a_{+, {\bf k}_1 + {\bf q} \sigma_1}^{\dagger}a_{-,{\bf k}_2 - {\bf q} \sigma_2}^{\dagger}
	a_{-, {\bf k}_2 \sigma_3}a_{+, {\bf k}_1 \sigma_4}.
	\label{Hamiltonian}
\end{align}
	The first term denotes the band energy with the q1d dispersion:
\begin{align}
	\xi ({\bf k})&= -2t_a \cos k_a -2t_b \cos k_b -\mu \nonumber\\
	&\simeq v_F (|k_a | -k_F^0 ) -2t_b \cos k_b + 2t_b' \cos 2k_b ,
	\label{dispersion}
\end{align}
	where the {\it intra}chain electron spectrum with the nearest {\it intra}chain hopping
	$t_a $ is 
	linearized at the Fermi points
	$\pm k_F^0 $ with the Fermi velocity $v_F$, and $\mu$ is the chemical potential.
	In Eq. (\ref{Hamiltonian}), $a_{p{\bf k} \sigma }^{\dagger}$ ($a_{p{\bf k} \sigma}$)
	is a creation (an annihilation) operator close to 
	the right $ k_a =+k_F^0 $ $(p=+)$ and the left $k_a = -k_F^0 $ $(p=-)$ Fermi surface, respectively.
	The second and third terms of Eq. (\ref{dispersion}) denote
	the nearest {\it inter}chain hopping $t_b$
	and the next-nearest {\it inter}chain hopping $t_b'$, respectively,
	where
\begin{align}
	t_b' =\frac{t_b ^2 \cos (k_F^0 )}{4t_a \sin^2 (k_F^0 )},
\end{align}
	is obtained as the nesting deviation.
	For the interaction part, the second term of Eq. (\ref{Hamiltonian}), 
	we follow the standard notations of the g-ology,
	where $g_1$ corresponds to backward scattering and $g_2$ does to forward scattering.
	(Note that $g_{1, 2}$ are normalized by $\pi v_F $ here.)
	For the 1/4-filling case, 
	$g_1 $ and $g_2$ are given in terms of on-site $U$, nearest-neighbor $V_1$ 
	and next-nearest-neighbor $V_2$ interactions, as
\begin{align}
	\pi v_F g_1 &= U/2 - V_2 , \\
	\pi v_F g_2 &= U/2 +V_1 +V_2 .
\end{align}
	In order to examine superconductivity and density wave
	in terms of RG technique,
	we utilize the path integral method\cite{DB},
	where the partition function $Z$ ($={\rm Tr}e ^{-\beta H}$) is given by
\begin{align}
	Z = \int \!\!\!\int \mathcal{D}\psi^* \mathcal{D}\psi
	\exp (S[\psi^* , \psi ]_\ell ),
\end{align}
	and the effective action $S$
	is written in the form,
\begin{align}
	S[\psi^* , \psi ]_\ell &=S_0 [\psi^* , \psi ]_\ell 
	+ S_{\rm I} [\psi^* , \psi ]_\ell , 
	\label{action}\\
	S_0 [\psi^* , \psi ]_\ell 
	&=\sum_{p, \sigma , \{ \tilde{{\bf k}}\} ^*}[G_p^0 (\tilde{{\bf k}})]^{-1}
	\psi_{p, \sigma}^* (\tilde{{\bf k}})\psi_{p, \sigma} (\tilde{{\bf k}}), \\
	S_{\rm I} [\psi^* , \psi ]_\ell &= 
	-\pi v_F \sum_{\mu, \tilde{\bf Q}}J_{\mu}(k_{1b}, k_{2b}, q_b ; \ell )
	O_{\mu}^{{\rm P}*} (\tilde{\bf Q})O_{\mu}^{\rm P} (\tilde{\bf Q}), \label{Js}\\
	&= -\pi v_F \sum_{\mu, \tilde{\bf q}}W_{\mu}(k_{1b}, k_{2b}, q_b ; \ell )
	O_{\bar{\mu}}^{{\rm C}*} (\tilde{\bf q})O_{\bar{\mu}}^{\rm C} (\tilde{\bf q}),
	\label{Ws}
\end{align}
	where the renormalization of 
	the bandwidth is defined as 
\begin{align}
	E_0 (\ell ) = E_0 \exp (-\ell ),
\end{align}
	with the bandwidth cut-off $E_0$.
	The quantities $\psi (\tilde{\bf k})$ and $\psi^* (\tilde{\bf k})$
	are the Grassmann fields
	and $\tilde{\bf k}=({\bf k}, \omega_n )$.
	$\{ {\bf k}\}^* $ denotes the renormalized momentum space.
	The operator $O_{\mu}^{{\rm P}}$ denotes Peierls fields
\begin{align}
	O_{\mu}^{{\rm P}}(\tilde{\bf Q})=
	\sqrt{\frac{T}{LN_b}}\sum_{\alpha , \beta , \{ \tilde{\bf k}\} ^*}
	\psi_{-, \alpha}^* (\tilde{\bf k}-\tilde{\bf Q})
	\sigma_{\mu}^{\alpha \beta}
	\psi_{+, \beta} (\tilde{\bf k}),
\end{align}
	describing the CDW ($\mu = 0$) and the SDW ($\mu = 1, 2, 3$) channels,
	where $\tilde{\bf Q}= (2k_F^0 + Q_0 , Q_b , \omega_m = 0, \pm 2\pi T, \cdots)$
	($Q_0\ll 2k_F^0$).
	The operator $O_{\bar{\mu}}^{{\rm C}}$ is Cooper fields
\begin{align}
	O_{\mu}^{{\rm C}}(\tilde{\bf q})=
	\sqrt{\frac{T}{LN_b}}\sum_{\alpha , \beta , \{ \tilde{\bf k}\} ^*}
	\alpha \psi_{-, -\alpha}^* (-\tilde{\bf k}+\tilde{\bf q})
	\sigma_{\bar{\mu}}^{\alpha \beta}
	\psi_{+, \beta} (\tilde{\bf k}),
\end{align}
	for singlet ($\bar{\mu}=0$) and triplet ($\bar{\mu}=1, 2, 3$) pairing,
	where $\tilde{\bf q}= (q_0 , q_b , \omega_m )$
	($q_0\ll 2k_F^0$).
	The couplings $J_{\mu}$ and $W_{\bar{\mu}}$ 
	correspond to that of the density wave and the superconductivity respectively,
	and are
	related to $g_{1, 2}$ as
	$J_{\mu =0}=g_1 -g_2 /2$, 
	$J_{\mu \neq 0}= -g_2 /2$, 
	$W_{\bar{\mu} =0}=-(g_1 + g_2 )/2$ and
	$W_{\bar{\mu} \neq 0}=(g_1 -g_2 )/2$, respectively.
%
	
%
	
%

\subsection{Renormalization of couplings $J_{\mu }(\ell )$, $W_{\bar{\mu }}(\ell )$}
	First we apply the RG technique to coupling constants for 
	the Peierls and the Cooper channel.
	By taking contraction of each channel, the one-loop flow equation 
	are obtained as\cite{DB}
\begin{align}
	\frac{d}{d\ell}J_{\mu}& (Q_b -k_b , k_b ; \ell ) \nonumber \\
	&= \frac{1}{N_b}\sum _{\bar{\mu}, k_b'}
	c_{\mu , \bar{\mu}} W_{\bar{\mu}}(Q_b -k_b , k_b' ; \ell ) 
	 W_{\bar{\mu}}(k_b' , k_b ; \ell ) I_{\rm C}(\ell ) \nonumber \\
	&-J_{\mu} (Q_b -k_b , k_b ; \ell )\frac{1}{N_b}\sum _{k_b'}
	J_{\mu} (Q_b -k_b' , k_b' ; \ell )I_{\rm P}(Q_b , k_b' ; \ell ),
	\label{q1dflow}
\end{align}
	where $c_{0,0}=-1/2, c_{0, \bar{\mu}\neq 0}=1/2, c_{\mu \neq 0, 0}=1/2$ 
	and $c_{\mu \neq 0, \bar{\mu}\neq 0}=1/6$.
	The first term of the right-hand side of Eq. (\ref{q1dflow})
	corresponds to the contraction in the Cooper channel,
	where $I_{\rm C} (\ell ) d\ell =\tanh [ E_0 (\ell )/4T] d\ell$
	is the outer-shell integration.
	The second term corresponds to the contraction in the Peierls channel,
	with the outer-shell integration
\begin{align}
	I_{\rm P}(Q_b , k_{b1}, \ell ) d\ell &=\frac{E_0 (\ell )d \ell}{4}\sum_{\lambda = \pm 1}
	\frac{1}{E_0 (\ell ) + \lambda A(k_{b1}, Q_b )} \nonumber \\
	&\times \biggl[ \tanh \frac{E_0 (\ell )}{4T} 
	+\tanh \frac{1}{2T}\bigl\{ \frac{E_0 (\ell )}{2}
	+\lambda A(k_{b1}, Q_b ) \bigr\} \biggr], \\
	A(k_b , Q_b )&=2t_b [ \cos k_b + \cos (k_b + Q_b ) ] \nonumber \\
					&+ 2t_b ' [ \cos 2k_b + \cos 2(k_b + Q_b )].
\end{align}
	For $t_b = 0$, 
	the flow equation (\ref{q1dflow}) yields the one-loop 
	1d flow equations.\cite{BC,Solyom}	
	Note that $I_{\rm C}(\ell )$ is a constant while $I_{\rm P}(Q_b , k_{b}, \ell )$
	suppressed by $t_b '$ depending on the wave vector of particles.
	Thus the Cooper channel is always singular 
	(i.e., does not depend on the shape of the Fermi surface)
	while the Peierls channel is suppressed by the nesting deviation.
	The renormalized Cooper couplings are obtained from
	the renormalized Peierls couplings through the relation
\begin{align}
	W_{\bar{\mu}=0}(Q_b - k_{b}, k_b ; \ell )&= 
	-\frac{1}{2}J_0 (Q_b - k_{b}, k_b ; \ell ) \nonumber \\
	&\hspace{0.5cm}+ \frac{3}{2}J_{\mu \neq 0}(Q_b - k_{b}, k_b ; \ell ), \label{WsJ}\\
	W_{\bar{\mu}\neq 0}(Q_b - k_{b}, k_b ; \ell )
	&= \frac{1}{2}J_0 (Q_b - k_{b}, k_b ; \ell ) \nonumber \\
	&\hspace{0.5cm}+ \frac{1}{2}J_{\mu \neq 0}(Q_b - k_{b}, k_b ; \ell ), \label{WtJ}
\end{align}
	which is obtained from Eqs. (\ref{Js}) and (\ref{Ws}) straightforwardly.

	Following the idea of the dimensional crossover theory\cite{BC,KY},
	we adopt two-step RG procedure.
	First we apply the RG technique of 1d\cite{BC} (1d RG) to the regime
	above the crossover temperature $T_{\rm cross}$,
	and then that of q1d\cite{DB} (q1d RG) to the region below $T_{\rm cross}$ 
	as the second step.
	For the first step, we adopt
	the two-loop flow equations of 1d\cite{BC}
\begin{align}
	\frac{d g_1 (\ell)}{d\ell}&=-g_1 (\ell )^2 -\frac{1}{2}g_1 (\ell )^3, \\
	\frac{dG(\ell )}{d\ell} &=0,
\end{align}
	where $G(\ell )\equiv g_1 (\ell ) -2g_2 (\ell )$.
	During the 1d RG, the {\it inter}chain coupling $j_{\mu}$ due to 
	the pair hopping is produced as follows:
\begin{align}
	j_{\mu }(\ell )&= z(\ell )^{-2}f_{\mu } (\ell ), \\
	\frac{d f_{\mu=0} (\ell )}{d \ell } &= \biggl[ 
	\frac{(g_2 (\ell ) -2g_1 (\ell ) ) z(\ell )t_b }{E_0 (\ell )} \biggr] ^2 , \\
	\frac{d f_{\mu \neq 0} (\ell )}{d \ell } &= \biggl[ 
	\frac{g_2 (\ell ) z(\ell )t_b }{E_0 (\ell )} \biggr] ^2 ,
\end{align}
	where flow equations for $t_b $ and $z$ 
	(corresponding to self energy) are 
\begin{align}
	\frac{d}{d\ell}\ln t_b (\ell )&=
	1-\frac{1}{4}\bigl[ g_1 (\ell )^2 + g_2 (\ell )^2 -g_1 (\ell )g_2 (\ell )
	\bigr] , \\
	\frac{d}{d\ell}\ln z (\ell )&=
	-\frac{1}{4}\bigl[ g_1 (\ell )^2 + g_2 (\ell )^2 -g_1 (\ell )g_2 (\ell ) \bigr] ,
\end{align}
	respectively.
	The dimensional crossover temperature is given
	by $T_{\rm cross}=E_0 (\ell_{\rm cross})/2$, 
	where $\ell _{\rm cross}$ is defined as 
	$E_{0}(\ell_{\rm cross})=t_b (\ell_{\rm cross})$.
	Thus the initial couplings for the second step, i.e., q1d RG,
	are given by a combination of $g_{\mu}$ and $j_{\mu} \cos Q_b$ as
\begin{align}
	J_{\mu =0} (Q_b , \ell_{\rm cross}) &=-\frac{1}{2}\bigl[ g_2 (\ell_{\rm cross})
	-2g_1 (\ell_{\rm cross}) \bigr] + j_{0} \cos Q_b ,\\
	J_{\mu \neq 0} (Q_b , \ell_{\rm cross}) &=-\frac{1}{2} g_2 (\ell_{\rm cross})
	 + j_{\mu} \cos Q_b .
\end{align}
%

\subsection{Response functions $\chi $}
	In order to calculate the response function,
	we add the source field to Eq. (\ref{action}),
	which is given by\cite{DB} 
\begin{align}
	S_h [\psi^* , \psi ]
	&=\sum_{\mu} z_{\mu} [O_{\mu}^{\rm P *}({\bf Q}_{\rm P})h_{\mu}({\bf Q}_{\rm P})
	+ {\rm c. c}] \nonumber \\
	& +\sum_{\bar{\mu}, \{ \tilde{\bf k}\} ^* }z_{\bar{\mu}}^{(n)} 
	[O_{\bar{\mu}}^{\rm C *}(\tilde{\bf k})h_{\bar{\mu}}^{(n)}(k_b )
	+ {\rm c. c}] ,
\end{align}
	where ${\bf Q}_{\rm P}=(2k_F^0 , \pi)$,
	and $h_{\bar{\mu}}^{(n)}(k_b )=h_{\bar{\mu}}^{(0)}\cos (nk_b )$
	or $h_{\bar{\mu}}^{(n)}(k_b )=h_{\bar{\mu}}^{(0)}\sin (nk_b )$.
	The corresponding pairing symmetries are discussed later.
	Here $z_{\mu (\bar{\mu})}^{(n)}$ corresponds to vertex corrections,
	and the value $[z_{\mu (\bar{\mu})}^{(n)}]^2$ is equal to the well-known auxiliary
	susceptibility $\bar{\chi}$.\cite{Solyom}
	Within the linear response theory,
	only the linear and the quadratic term contribute to the action,
	so that the action at $\ell$ has the form
\begin{align}
	S[\psi ^* , \psi , h^* , h]_{\ell} &= S[\psi ^* , \psi ]_{\ell}
	+\sum_{\mu} z_{\mu}(\ell ) [O_{\mu}^{\rm P *}({\bf Q}_{\rm P})h_{\mu}({\bf Q}_{\rm P})
	+ {\rm c. c}] \nonumber \\
	&\hspace{-15mm} +\sum_{\bar{\mu}, \{ \tilde{\bf k}\} ^* }z_{\bar{\mu}}^{(n)}(\ell )
	[O_{\bar{\mu}}^{\rm C *}(\tilde{\bf k})h_{\bar{\mu}}^{(n)}(k_b )
	+ {\rm c. c}] \nonumber \\
	&\hspace{-15mm}+\sum_{\mu} \chi _{\mu}(\ell )h_{\mu}^*({\bf Q}_{\rm P})h_{\mu}({\bf Q}_{\rm P})
	+\sum_{\bar{\mu}, k_b }
	\chi_{\bar{\mu}}^{(n)}(\ell )h_{\bar{\mu}}^{(n) *} (k_b )h_{\bar{\mu}}^{(n)} (k_b ).
\end{align}
	Flow equations of each $z_{\mu (\bar{\mu})}^{(n)}$ are given by
\begin{align}
	\frac{d}{d\ell}\ln z_{\mu}(\ell )=-\frac{1}{N_b}\sum_{k_b}
	J_{\mu } (\pi -k_b , k_b ; \ell ) I_{\rm P}(\pi , k_b ; \ell ), 
\end{align}
	for the Peierls channel and 
\begin{subequations}
\begin{align}
	\frac{d}{d\ell} \ln z_{\bar{\mu}=0}^{(n)}(\ell )
	&= \frac{1}{2}c_{\bar{\mu}=0}^{(n)} (\ell ) I_{\rm C}(\ell ), \\
	\frac{d}{d\ell}\ln z_{\bar{\mu}\neq 0}^{(n)}(\ell )
	&= \frac{1}{2}c_{\bar{\mu}\neq 0}^{(n)} (\ell ) I_{\rm C}(\ell ),
\end{align}
\end{subequations}
	for the Cooper channel.
	Then the static response functions are obtained as
\begin{align}
	\chi _{\mu (\bar{\mu})}^{(n)}=(\pi v_F )^{-1}
	\int_0^{\ell} \bigl[ z_{\mu (\bar{\mu})}^{(n)} (\ell ')\bigr] ^2 d\ell '.
	\label{response}
\end{align}
	The coefficients $c_{\bar{\mu}}^{(n)}$ represent the Fourier coefficients
	of the coupling in Cooper channel $a_{\bar{\mu}}^{(n)}$
	and $b_{\bar{\mu}}^{(n)}$, which are defined as 
\begin{align}
	W_{\bar{\mu}}(k_b , k_b ' ; \ell )
	=&a_{\bar{\mu}}^{(0)} (\ell )
	+\sum_{n>0}\bigl[ 
	a_{\bar{\mu}}^{(n)} (\ell )\cos (nk_b ) \cos (nk_b ' ) \nonumber \\
	&+b_{\bar{\mu}}^{(n)} (\ell )\sin (nk_b ) \sin (nk_b ' ) \bigr] .
\end{align}
%
%
	The coefficients $a_{\bar{\mu}}^{(n)}$ ($b_{\bar{\mu}}^{(n)}$) are the even 
	(odd) components with respect to the {\it inter}chain direction $k_b , k_b'$.
	The parity of the gap cannot be classified only by $k_b$-dependence.
	It has to be classified by the symmetry of both $k_a$- and 
	$k_b$-dependences of the gap function.
	For singlet channel, considering only the even-frequency pairing, 
	the parity of the gap is even,
	so that 
	\begin{description}
		\item[\hspace{10mm}A)] $a_{\bar{\mu}=0}^{(n)}$: even-$k_a$ and even-$k_b$ 
		($s$, $d_{x^2 -y^2}$, ...)
		\item[\hspace{10mm}B)] $b_{\bar{\mu}=0}^{(n)}$: odd-$k_a$ and odd-$k_b$
		($d_{xy}$, $g$, ...)
	\end{description}
	are possible.
	Similarly for triplet channel, the parity of the gap is odd,
	so that 
	\begin{description}
		\item[\hspace{10mm}C)] $a_{\bar{\mu}\neq 0}^{(n)}$: odd-$k_a$ and even-$k_b$ 
		($p_x$, $f_x$, ...)
		\item[\hspace{10mm}D)] $b_{\bar{\mu}\neq 0}^{(n)}$: even-$k_a$ and odd-$k_b$
		($p_y$, $f_y$, ...)
	\end{description}
	are possible.
	Briefly speaking, the symmetry of $k_b$-dependence is classified
	by $a_{\bar{\mu}}^{(n)}$ or $b_{\bar{\mu}}^{(n)}$,
	and that of
	$k_a$-dependence is automatically determined once we choose
	the spin-symmetry $\bar{\mu}$.

\begin{figure}[t]
\begin{center}
\includegraphics[width=8cm]{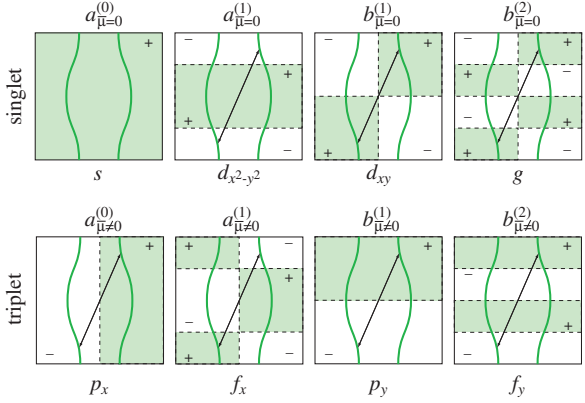}
\end{center}
\caption{
	Types of the gap symmetry. The upper panels corresponds to the singlet
	pairing, whose symmetry belongs to $s$-, $d_{x^2 -y^2}$-, 
	$d_{xy}$-, 	$g$-wave respectively (from left to right).
	The lower panels corresponds to the triplet
	pairing, whose symmetry belongs to $p_x$-, $f_x$-, $p_{y}$-, 
	$f_y$-wave 	respectively (from left to right).
	The coefficients ($a_{\bar{\mu}}^{(n)}, b_{\bar{\mu}}^{(n)}$) above the panels denote
	the corresponding Fourier coefficients.
	The arrows indicate the inversion (${\bf k} \to -{\bf k}$), 
	which does not (does) change the sign of 
	the gap function for singlet (triplet) pairing.
	}
\label{spdfg}
\end{figure}

	In this paper, we consider the following 8 symmetries
	for order parameters of superconductivity.
	For singlet channel, we calculate the response functions for
	$a_{\bar{\mu}=0}^{(0)}$, $a_{\bar{\mu}=0}^{(1)}$, $b_{\bar{\mu}=0}^{(1)}$
	and $b_{\bar{\mu}=0}^{(2)}$, which can be recognized as 
	$s$-, $d_{x^2 -y^2}$-, $d_{xy}$- and $g$-wave, respectively.\cite{symmetry}
	For triplet channel, the response functions for $a_{\bar{\mu}\neq 0}^{(0)}$, 
	$a_{\bar{\mu}\neq 0}^{(1)}$,
	$b_{\bar{\mu}\neq 0}^{(1)}$
	and $b_{\bar{\mu}\neq 0}^{(2)}$, which can be recognized as 
	$p$-, $f_x$-, $p_y$- and $f_y$-wave respectively,
	are calculated.
	(See Fig. \ref{spdfg} for the graphical representation of 
	each symmetry.)
	We examined whole 8 symmetries,
	and found the clear instability only for
	$p_x$-, $d_{x^2 -y^2}$- and $f_x$-wave.
	Therefore, from now on, we simply call them $p$-, $d$-, $f$-wave, respectively.

\section{Superconductivity vs. Density Wave}
\subsection{Temperature dependence of $\chi (T)$ and $T_{\rm c}$}

\begin{figure}[htbp]
\begin{center}
\includegraphics[width=8cm]{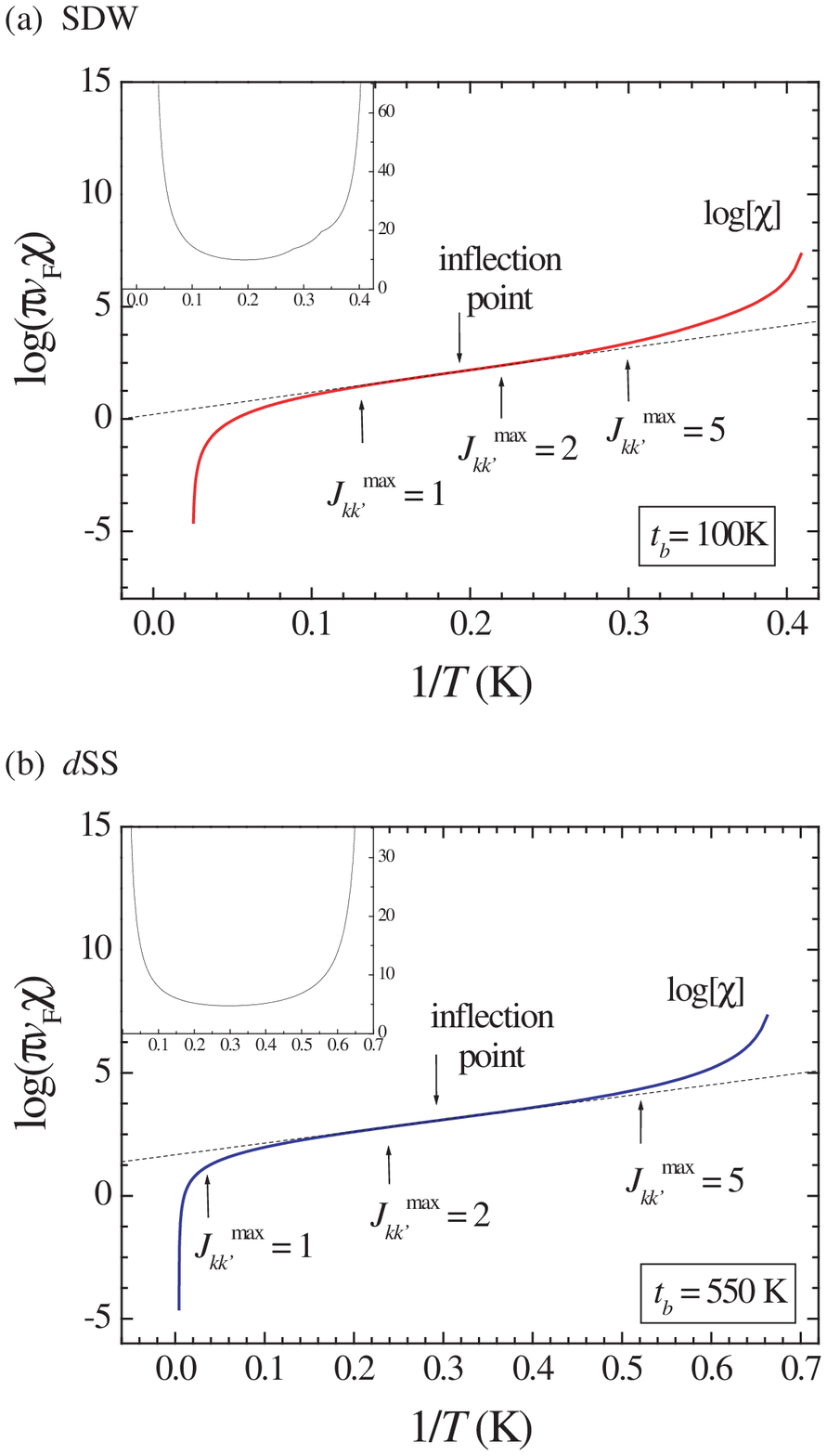}
\end{center}
\caption{Response function of (a) SDW and (b) $d$SS.
The dotted straight line is the line of $\tilde{\chi}_{\rm prop}$
drawn by fitting with $\tilde{\chi}$
at the inflection point.
The insets show $d (\ln [\pi v_F \chi ])/d(1/T)$ for the respective response function.}
\label{res}
\end{figure}
	Based on the RG method derived in previous section,
	we examine the possible states in the present q1d system.
	First we calculate the response functions obtained from Eq. (\ref{response}),
	which are 
	shown in Fig. \ref{res} for the SDW (a) and the $d$SS (b).
	Each response function (also for other response functions) shows
	the similarity in behavior.
	The important point to note is that the response function $\chi (T)$ 
	becomes proportional
	to $\exp (1/T)$ (i.e., $\ln \chi (T) \propto 1/T $)
	in the wide region of intermediate temperatures.
	From the theorems of Mermin and Wagner\cite{MW},
	and Hohenberg\cite{Hohenberg},
	there is 
	no magnetic or superconducting long-range order 
	at any finite temperature in one and two dimensions.
	Generally, the response functions in two dimension have the temperature dependence
	$\chi (T) \propto \exp (1/T)$ for both magnetic\cite{KT}
	and superconducting\cite{TMY,private,Kosuge} cases,
	when we do not consider the Kosterlitz-Thouless transition.\cite{KT2}
	From these points, one may say that
	the present method gives a rigorous solution in such temperatures
	as $\chi (T) \propto \exp (1/T)$
	in spite of the one-loop approximation.

	In the very low temperatures, on the other hand,
	$\chi (T)$ strays from the behavior of $\exp (1/T)$
	and diverges at finite temperature
	because of the divergence of the coupling $J$ or $W$.
	Therefore, it is appropriate to recognize 
	the response function in the intermediate temperatures
	as a proper response function, since it is proportional to $\exp (1/T)$.
	In the region, where the obtained response function 
	does not show the behavior of $\exp (1/T)$, the present RG technique 
	does not work well due to the relevant (large enough) couplings.
	In the sense, the characteristic temperature, $T^*$,
	for the lower bound of the present RG technique, 
	may be taken by a condition that 
	$d (\ln [\pi v_F \chi ])/dT^{-1}$ takes a minimum, i.e., the inflection point.
	In order to obtain the proper response function,
	we estimate by using an extrapolation formula
\begin{align}
	\tilde{\chi}_{\rm prop} (T) =\alpha \exp ^{\beta /T},
\end{align}
	where $\tilde{\chi}=\pi v_F \chi$.
	The parameter $\beta$ 
	is defined as $\beta \equiv d (\ln [\tilde{\chi} ])/dT^{-1}|_{T=T^*}$,
	which is of the order of the transition temperature
	obtained by the mean field approximation.
	The parameter $\alpha$ is defined as
	$\alpha \equiv \tilde{\chi} (T^*)e^{-\beta /T^*}$,
	namely, $\tilde{\chi}_{\rm prop} (T)$ is the tangential line of 
	$\ln \tilde{\chi}(1/T)$ at $T=T^*$.
	Assuming that the practical transition temperature $T_{\rm c}$ is lead by
	the three-dimensionality,
	we use the conventional RPA (for the inter-plane coupling),
	$g_{\rm 3d}\tilde{\chi}_{\rm prop} (T_{\rm c})=1$,
	which yields the form 
\begin{align}
	T_{\rm c} = -\frac{\beta}{\ln (\alpha g_{\rm 3d})},
	\label{Tccondition}
\end{align}
	where $g_{\rm 3d}$ is an inter-plane coupling.
	Of course, we need to discuss
	the problem in the Kosterlitz-Thouless (KT) context
	in two-dimensional systems.\cite{KT2}
	In the case of $T_{\rm c}\ll E_F$, however,
	the difference between $T_{\rm c}$ and 
	the KT transition temperature $T_{\rm c}^{\rm KT}$ is small
	compared to $T_{\rm c}$, i.e., $|T_{\rm c}-T_{\rm c}^{\rm KT}|/T_{\rm c}
	\sim T_{\rm c}/E_F$.\cite{Miyake}
	Consequently, $T_{\rm c}$ determined by the present method
	almost corresponds to the actual transition temperature $T_{\rm c}^{\rm KT}$.

	In this paper, $T_{\rm c}$ is calculated by taking
	$\mathcal{O}(g_{\rm 3d}) \sim 10^{-3}$.
	We calculated $32\times 32\times 2$ simultaneous differential equations,
	i.e., a 32-chains system with 64 points on the Fermi surface.
	The initial bandwidth is set to be $E_0 =5500K$ 
	($t_a = E_0/(2\sqrt{2}k_F^0)\simeq 2500$K ),
	which is consistent with the bandwidth of (TMTSF)$_2$PF$_6$.\cite{JTB,Grant}

\subsection{$g_2$-$g_1$ phase diagram}
\begin{figure}[t]
\begin{center}
\includegraphics[width=9cm]{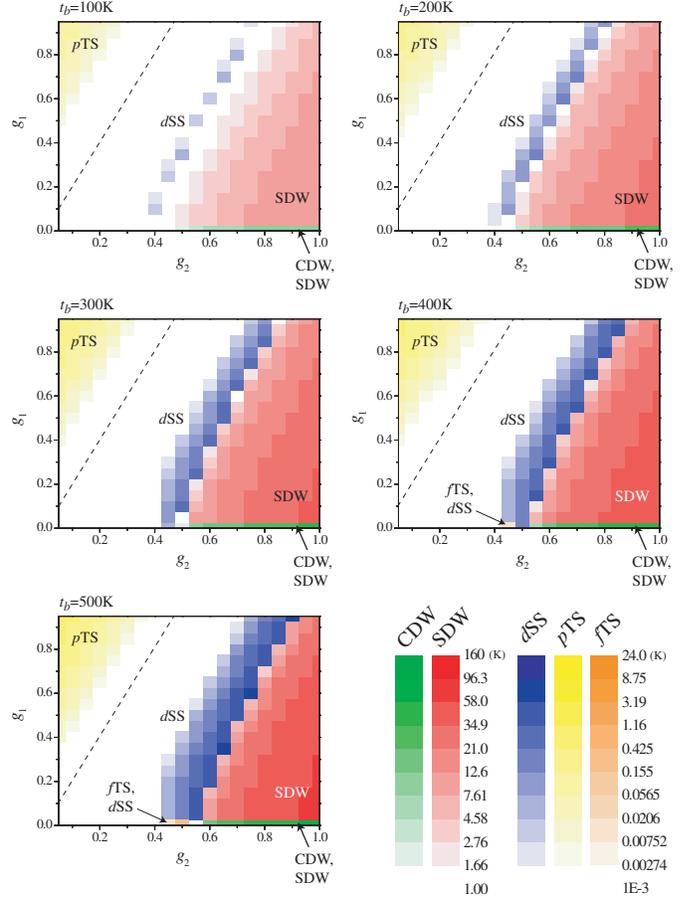}
\end{center}
\caption{
$g_2$-$g_1$ phase diagrams for $t_b =100, 200, 300, 400$ and 500K.
The transition temperatures for the CDW, SDW, $d$SS, $p$TS and $f$TS
are expressed as the tones of green, red, blue, yellow and orange, respectively.
The dashed-line indicates the phase boundary between the $p$TS and the SDW
of the g-ology.
}
\label{g1g2}
\end{figure}

	First, we show phase diagrams on the $g_2$-$g_1$ plane for several choices of 
	$t_b$'s
	in order to understand the variation of the phase diagram
	from pure 1d case to q1d case.
	For repulsive interactions $g_{1, 2}>0$,
	the SDW is dominant for $g_1 <2g_2$
	and the $p$-wave triplet superconductivity ($p$TS)
	is dominant for $g_1 > 2g_2$ as found in the g-ology.\cite{Solyom}
	Figure \ref{g1g2} shows the phase diagram for $t_b =$ 100, 200, 300, 400, 500K.
	The dashed-lines in Fig. \ref{g1g2} denote $g_1$=$2g_2$, 
	which correspond to the phase boundary between the $p$TS and the SDW
	of the g-ology.

	The present results show that the transition temperatures of the $p$TS and the SDW
	are extremely low (numerically, below $10^{-2}$ K) near the phase boundary $g_1 =2g_2$.
	Moving from that line, 
	the $p$TS and the SDW phase appears parallel to the $g_1$=$2g_2$-line.
	The salient feature is
	that, in q1d, the $d$SS phase emerges between the $g_1$=$2g_2$-line and the SDW phase,
	and this $d$SS phase extends to the SDW phase
	also parallel to $g_1$=$2g_2$,
	as $t_b$ increases.
	From these features,
	it turns out that the transition temperature 
	depends on the intensity of $|g_1 -2g_2|$,
	corresponding to the intensity of density fluctuations.
	The nearer the parameters move to the line of $g_1$=$2g_2$, 
	the weaker the SDW instability becomes.
	Thus the $d$SS appears
	instead of the SDW state due to the deviation of the 
	nesting condition.
	Just on the line of $g_1$=0, the CDW ($f$TS) state
	has the same transition temperature as the SDW ($d$SS) state.
	The phase boundary between the $d$SS and the normal state 
	swerves from the line parallel to $g_1 =2g_2$ for small $g_1$,
	while that between the $d$SS and the SDW is parallel to the $g_1$=$2g_1$-line.
	This can be understood as the interference effect 
	of the charge fluctuations and the spin fluctuations,
	which shall be discussed later.

\subsection{$t_b$-dependence of transition temperature}

\begin{figure}[tb]
\begin{center}
\includegraphics[width=8cm]{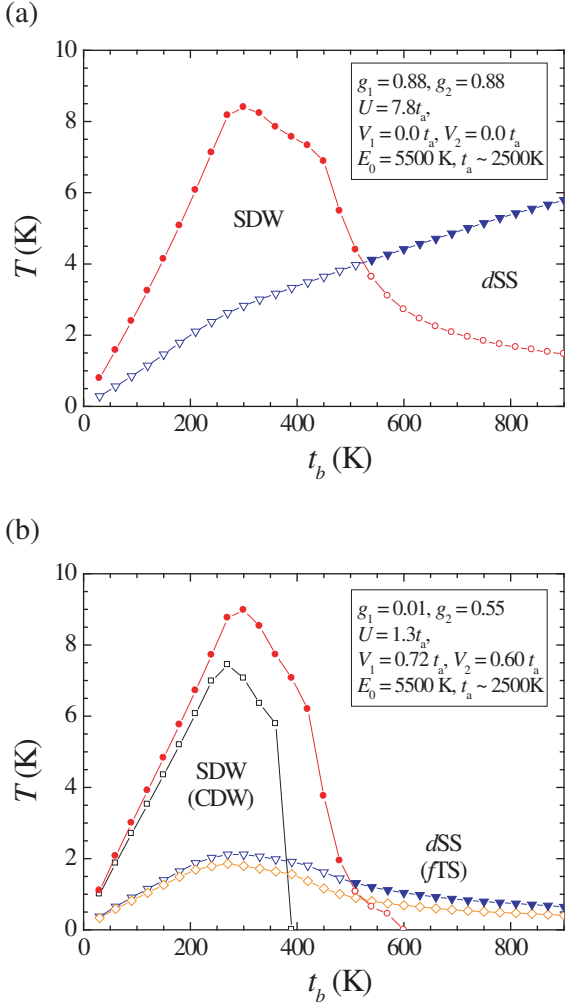}
\end{center}
\caption{Transition temperature as a function of the {\it inter}chain hopping $t_b$
for (a) the Hubbard ($V$=0) model $g_1 =g_2 =0.88$ ($U \simeq 7.8 t_a , V_1 = V_2 =0.0$),
and for (b) the large-$V$ model $g_1 = 0.01, g_2 = 0.55$ ($U \simeq 1.3 t_a,
V_1 \simeq 0.72 t_a , V_2 \simeq 0.60 t_a $).
The closed circles and triangles are the transition temperature of SDW and $d$SS,
respectively. The open circles and diamonds indicates the temperature
where the response functions of CDW and $f$TS satisfy the condition
(\ref{Tccondition}), respectively.}
\label{Ttb}
\end{figure}

	Next, we show $t_b$-dependence of the transition temperature 
	in Fig. \ref{Ttb}; 
	(a) the Hubbard ($V$=0) model with
	$g_1 =g_2 =0.88$ ($U \simeq 7.8 t_a , V_1 = V_2 =0.0$),
	and (b) the extended Hubbard model with large-$V$ given by
	$g_1 = 0.01, g_2 = 0.55$ ($U \simeq 1.3 t_a,
	V_1 \simeq 0.72 t_a$ and $V_2 \simeq 0.60 t_a $, if we assume 
	$V_1 = 1.2V_2$).	
	For small $t_b$ ($\lesssim $300K), the transition temperature
	of SDW, $T_{\rm SDW}$, increases linearly.
	More precisely,
	it is approximately given by $T_{\rm SDW}\sim t_b \exp (-2/g_2^*)$.
	Here, $g_2^* /2 $ corresponds to the strength of the spin fluctuation,
	where the asterisk denotes its value at $T_{\rm cross}$.
	Such increase cannot be obtained by the mean field theory
	or the RPA-type theory,
	and consistent with the previous 1d RG theory;\cite{SF,BC}
	this feature originates from the effect of 1d fluctuation,
	i.e., the equally divergent contributions from 
	both the Cooper and the Peierls channel.
	The transition temperature has a maximum followed by the decrease
	due to the nesting deviation.
%
%
	When $T_{\rm SDW}$ is sufficiently suppressed due to the nesting deviation,
	the {\it hidden} superconductivity, $d$SS in the present case, emerges.
	These features are seen for almost every couplings in the parameter region
	$0<g_1 < 2g_2$, corresponding to the SDW phase of the g-ology.

	There are some differences between the $V$=0
	and the large-$V$ model.
	The most noticeable difference exists in the subdominant phase 
	for the large-$V$,
	which is shown by open squares (CDW) and diamonds ($f$TS) in Fig. \ref{Ttb} (b).
	Here, the transition temperature for the subdominant state
	are also obtained from the condition (\ref{Tccondition}).
	The transition temperature for the subdominant CDW ($f$TS)
	is exactly the same as that of SDW ($d$SS) when
	$g_1 =0$ ($U=2V_2$), i.e., just on the phase boundary between
	SDW and CDW of the g-ology.
	The actual phase transition, of course, occurs 
	only for the most divergent response function,
	so that these subdominant phases are masked practically.
	Therefore the subdominant transition temperature
	is better considered as an emergence of fluctuations.
	In the case of large-$V$,
	not only the spin fluctuations but also the charge fluctuations
	develop, while only the spin fluctuations 
	develop for the $V$=0.
	This enhancement of the charge fluctuation promotes the $f$TS state;
	especially, $f$TS is realized for $g_1 = 0$.

	Another significant difference is that the superconducting 
	transition temperature $T_{\rm c}$ for large-$V$ is much lower 
	than that for $V$=0,
	while $T_{\rm SDW}$ of both cases are almost the same.
	This comes from the difference of the strength of the charge fluctuations.
	It is clear from the eqs. (\ref{WsJ}) and (\ref{WtJ}) 
	that the charge fluctuation $J_0$ suppresses the singlet channel
	$W_{0}$, and enhances the triplet channel $W_{\bar{\mu}\neq 0}$.
	When $J_0 = J_{\mu \neq 0}$, the amplitude of the triplet channel
	becomes just the same as that of singlet one,
	which corresponds to the $g_1 =0$ ($U=2V_2$) case.
	Thus the low $T_{\rm c}$ of the large-$V$ model also suggests
	the enhancement of the charge fluctuation,
	which promotes the $f$TS.
	The $t_b$-dependence of $T_{\rm c}$ would be given by
	$T_{\rm c} \sim t_b \exp (-1/\tilde{g}_{\bar{\mu}}^*)$
	in the same way as the SDW.
	(Here, $\tilde{g}_{\bar{\mu}}^*$ is slightly smaller than $g_{\bar{\mu}}^*$
	because of the nodes of the gap.)
	Therefore, the increasing $T_{\rm c}$ for the Hubbard model
	is due to the increasing $t_b$.
	The reason why $T_{\rm c}$ decreases for the large-$V$
	is the rapid decrease of the spin fluctuation due to the cancellation 
	with the charge fluctuation.

	Note that the possible $f$TS state is a characteristic of 
	systems with open-Fermi-surface
	such as q1d.
	On the other hand, in the case of the closed Fermi surface, 
	such as the square lattice,
	$f$TS cannot have the comparable transition temperature
	even for large-$V$,
	but $d_{xy}$-wave singlet superconductivity can
	become a dominant state.\cite{OAKA,KTOS}
	This comes from a property of the open Fermi surface.
	The number of nodes of $f$TS is 4 in the case of q1d open Fermi surface,
	which is the same as $d$SS.
	Here, the number is 6 for the closed Fermi surface.
	Hence, the q1d system possesses a property which is favorable for 
	the triplet pairing.
	As far as the pairing symmetry is concerned,
	the present results is consistent with the RPA results.\cite{TK}
	The RPA, however, does not take into account the 1d fluctuations,
	so that its $T_{\rm c}$ is much higher than the present one.
%

\section{Conclusion}
	We have calculated the response functions and the transition temperatures 
	for the density wave (CDW and SDW) and 
	the superconductivity 
	($s$-, $p_{x}$-, $p_{y}$-, $d_{x^2 - y^2}$-, $d_{xy}$-, $f_x$-, $f_y$- and $g$-wave) 
	of q1d systems by the Kadanoff-Wilson renormalization group 
	technique, which has been developed by Duprat and Bourbonnais.\cite{DB}
	The present method is able to investigate a system in pure 1d and q1d
	in the same footing.
	Each response function exhibits the temperature dependence, 
	$\chi (T) \propto \exp (1/T)$,
	in a wide region of temperature, which 
	satisfies the theorems of the Hohenberg\cite{Hohenberg}, 
	and Mermin and Wagner.\cite{MW} 
	Also, the temperature dependence is consistent 
	with that obtained by 2d fluctuation theories.
	\cite{KT,TMY,private,Kosuge}
	This may suggest that the present method gives a rigorous solution
	to some extent.

	In the phase diagram of the $g_2$-$g_1$ plane,
	the $p$TS and the SDW are locating in the region parallel to $g_1$=$2g_2$.
	The transition temperature depends on the intensity $|g_1 -2g_2 |$,
	which corresponds to the coupling constant of density-fluctuations.
	The $d$SS phase, which does not appear in purely 1d case,
	exists next to the SDW phase in q1d case.
	The $d$SS phase expands into the SDW region parallel to $g_1$=$2g_2$
	as $t_b$ increases.
	For small $t_b$, $T_{\rm SDW}$ increases as 
	$T_{\rm SDW}\sim t_b \exp (-2/g_2^*)$.
	The transition temperature of the SDW has a maximum and decreases
	at finite $t_b$ ($\sim 300$K for $(g_1 , g_2 ) = (0.88, 0.88)$ and 
	$(g_1 , g_2 )=(0.01, 0.55)$), due to the nesting deviation.
	The singlet superconductivity of $d_{x^2 - y^2}$-wave ($d$SS) 
	emerges when the SDW state is suppressed
	by the moderately large $t_b$.
	In the large-$V$ model, the subdominant CDW ($f$-wave triplet) state
	is comparable to the dominant SDW ($d_{x^2 -y^2}$-wave singlet) state.
	Especially, the transition temperature of $f$TS (CDW)
	becomes equivalent to that of $d$SS (SDW)
	with $V_2 =U/2$.
	This is because the large $V$ (practically $V_2$) leads to 
	the enhancement of the charge fluctuations.
	Such large charge-fluctuation suppresses the $d$SS
	and enhances the $f$TS,
	so that $T_{\rm c}$ of $d$SS in the large-$V$ model 
	is much lower than that in the $V$=0 model.
	Consequently, the long-range Coulomb interaction 
	can give rise to the large charge fluctuation, which suppresses $d$SS
	and enhances $f$TS {\it particularly} in q1d systems.

	The experiments on (TMTSF)$_2$PF$_6$ compounds
	suggest the coexistence of $2k_F$-CDW and $2k_F$-SDW,\cite{PR,Kagoshima}
	and exhibit about ten times lower $T_{\rm c}$ than $T_{\rm SDW}$.\cite{Jerome}
	Considering these experimental properties,
	the large-$V$ model ($V_2 /U\sim 0.5$) is appropriate for the (TMTSF)$_2$PF$_6$ compounds
	on the point of
	the phase diagram, the $2k_F$-charge-fluctuation in SDW phase,
	and the realization of the $f$TS.
	Actually, there is an estimation that $V_1 /U \sim $ 0.5-0.75 
	($V_2 /U \sim$ 0.4-0.6, if we assume $V_1 = 1.2 V_2$)
	on a basis of a Valence-Bond/Hartree-Fock method for molecules
	derived of TTF.\cite{CFD}

	Even if $V_2$ is not so large,
	$f$TS is possibly realized under moderately 
	large Zeeman magnetic field,
	which suppresses the singlet pairing of $d$SS
	due to the paramagnetic effect\cite{FOKM}.
	In such a case, since the ground state
	is $d$SS state,
	the Knight shift decreases at $H=0$ but 
	does not vary under large magnetic field.
	Noting that the upper critical field can be estimated as 
	$H_{{\rm c}2}=(\Delta_0 /\sqrt{2} \mu _{\rm B})(1-T^2/T_{\rm c}^2 )$,\cite{Clogston}
	we can evaluate that
	1.7 T is sufficient to realize the field-induced $f$-wave triplet pairing
	for $T_{c}(H=0)=1.2$K, even for $T_{\rm c}^{f}\simeq T_{\rm c}^d /2$.\cite{Zeeman}
	Here, $\Delta_0$ is the magnitude of the superconducting gap,
	and the BCS relation $2\Delta_0 =3.5 k_{\rm B}T_{\rm c}$ is assumed.

	Finally, we comment on the effects of the dimerization
	and the Umklapp scattering, which are not considered in the present work.
	Following effect is expected with the Umklapp scattering $g_3$
	as $t_b$ increases.\cite{KY,BD}
	First the Umklapp scattering causes the antiferromagnetic transition
	($g_3$ is relevant while $g_1$ is irrelevant).
	Next, the SDW transition is realized
	due to the nesting property ($g_1$ becomes relevant).
	Finally, the $f$TS state results from the charge fluctuation 
	enhanced by the off-site interaction,
	which is not affected by the Umklapp scattering;
	the essential points of the present paper would be still valid.

	\section*{Acknowledgements}
	The authors thank M. Tsuchiizu for valuable discussions
	at the early stage of the present work.
	They are also grateful to
	C. Bourbonnais for a lot of useful comments.
	Y. F. acknowledges helpful discussions with K. Miyake
	on the two-dimensional fluctuations.
	The present work has been financially supported by 
	a Grant-in-Aid for Scientific Research on Priority Areas of
	Molecular Conductors (No. 15073103 and 15073213) from
	the Ministry of Education, Culture, Sports, Science and Technology, Japan.



\begin{thebibliography}{9}
\bibitem{symmetry}
	Of course, in q1d systems, we cannot classify
	the gap symmetry by the spherical-wave basis
	or the irreducible representation of the tetragonal.
	Here, we use an idiomatic representation
	such as $s$-, $p_{x, y}$-, $d_{x^2-y^2 , xy}$,
	$f_{x,y}$ and $g$ displayed in Fig. \ref{spdfg}.
\bibitem{Zeeman}
	The evaluated Zeeman magnetic field here
	is the magnetic field where
	the singlet pairings are completely broken.
	The reversal from $d$SS to $f$TS 
	is expected below the $H_{\rm c2}$ of the $d$SS.
\end{thebibliography}
\end{document}